\documentclass[conference]{IEEEtran}
\IEEEoverridecommandlockouts
\usepackage{multirow}
\usepackage{cite}
\usepackage{amsmath,amssymb,amsfonts}
\usepackage{footnote}
\usepackage{multirow}
\usepackage[flushleft]{threeparttable}
\usepackage{algorithmic}
\usepackage{graphicx}
\usepackage{textcomp}
\usepackage{xcolor}
\usepackage{hyperref}
\usepackage[acronym]{glossaries}
\newacronym{acm}{ACM}{automatic modulation classification}
\newacronym{acp}{ACP}{averaged cyclic periodogram}
\newacronym{aie}{AIE}{AI Engine}
\newacronym{ai}{AI}{artificial intelligence}
\newacronym{ar}{AR}{autoregressive}
\newacronym{arma}{ARMA}{autoregressive moving average}
\newacronym{agu}{AGU}{address generation unit}
\newacronym{bpsk}{BPSK}{binary phase-shift keying}
\newacronym{caf}{CAF}{cyclic autocorrelation function}
\newacronym{cdp}{CDP}{channel-data product}
\newacronym{cs}{CS}{cyclic spectrum}
\newacronym{cpu}{CPU}{central processing unit}
\newacronym{cms}{CMS}{cyclic modulation spectrum}
\newacronym{dit}{DIT}{decimation-in-time}
\newacronym{dif}{DIF}{decimation-in-frequency}
\newacronym{dft}{DFT}{discrete Fourier transform}
\newacronym{dsss}{DSSS}{direct-sequence spread-spectrum}
\newacronym{dsp}{DSP}{digital signal processing} 
\newacronym{ddrmc}{DDRMC}{DDR memory controller}
\newacronym{eda}{EDA}{electronic design automation}
\newacronym{fft}{FFT}{fast Fourier transform}
\newacronym{fam}{FAM}{FFT accumulation method}
\newacronym{facp}{FACP}{fast averaged cyclic periodogram}
\newacronym{fe}{FE}{frequency estimation}
\newacronym{fs}{FS}{frequency-smoothing}
\newacronym{fsc}{FSC}{fast spectral correlation}
\newacronym{fpga}{FPGA}{field-programmable gate array}
\newacronym{gpu}{GPU}{graphics processing unit}
\newacronym{hls}{HLS}{high-level synthesis}
\newacronym{ii}{II}{initiation interval}
\newacronym{lpf}{LPF}{low pass filter}
\newacronym{mac}{MAC}{multiply-accumulate}
\newacronym{ml}{ML}{machine learning}
\newacronym{noc}{NoC}{network-on-chip}
\newacronym{pca}{PCA}{principal component analysis}
\newacronym{psd}{PSD}{power spectral density}
\newacronym{pl}{PL}{programmable logic}
\newacronym{plio}{PLIO}{programmable logic input/output}
\newacronym{rfml}{RFML}{radio frequency machine learning}
\newacronym{rtl}{RTL}{register-transfer-level}
\newacronym{scd}{SCD}{spectral correlation density}
\newacronym{sdr}{SDR}{software-defined radio}
\newacronym{ssca}{SSCA}{strip spectral correlation analyser}
\newacronym{s3ca}{S$^3$CA}{sparse strip spectral correlation analyser}
\newacronym{sft}{SFFT}{sparse fast Fourier transform}
\newacronym{snr}{SNR}{signal-to-noise ratio}
\newacronym{sqnr}{SQNR}{signal-to-quantisation-noise ratio}
\newacronym{stft}{STFT}{short-time Fourier transform}
\newacronym{simd}{SIMD}{single-instruction multiple-data}
\newacronym{soc}{SoC}{system-on-chip}
\newacronym{ts}{TS}{time-smoothing}
\newacronym{tops}{TOPS}{tera operations per second}
\newacronym{tflops}{TFLOPS}{tera floating point operations per second}
\newacronym{vhdl}{VHDL}{VHSIC hardware description language}
\newacronym{vliw}{VLIW}{very-long instruction word}
\newacronym{2dssca}{SSCA\_2DFFT}{strip spectral correlation analyser utilising a decomposed FFT}
\newacronym{2dfft}{2DFFT}{decomposed FFT}

\usepackage{orcidlink}
\usepackage[ruled,vlined]{algorithm2e}
\SetKw{KwBy}{by}
\usepackage{subcaption}
\def\BibTeX{{\rm B\kern-.05em{\sc i\kern-.025em b}\kern-.08em
    T\kern-.1667em\lower.7ex\hbox{E}\kern-.125emX}}

\begin{document}

\title{AMD Versal Implementations of\\ FAM and SSCA Estimators \\
\thanks{The authors would like to thank the AMD/Xilinx University Program for the generous donation of a Versal VCK5000 board.}
}

\author{
Carol Jingyi Li$^{* \dagger}$\orcidlink{0000-0001-7638-6323}, Ruilin Wu$^{*}$\orcidlink{0009-0009-9927-0898}, Philip H.W. Leong$^{*}$\orcidlink{0000-0002-3923-3499}\\
$^{*}$ Computer Engineering Lab, The University of Sydney, NSW, Australia\\
$^{\dagger}$ Reconfigurable Computing Systems Lab
, The Hong Kong University of Science and Technology, Hong Kong\\
\{jingyi.li, ruilin.wu, philip.leong\}@sydney.edu.au
}

\maketitle

\begin{abstract}
Cyclostationary analysis is widely used in signal processing, particularly in the analysis of human-made signals, and spectral correlation density (SCD) is often used to characterise cyclostationarity. Unfortunately, for real-time applications, even utilising the fast Fourier transform (FFT), the high computational complexity associated with estimating the SCD limits its applicability. In this work, we present optimised, high-speed field-programmable gate array (FPGA) implementations of two SCD estimation techniques. 
Specifically, we present an implementation of the FFT accumulation method (FAM) running entirely on the AMD Versal AI engine (AIE) array. We also introduce an efficient implementation of the strip spectral correlation analyser (SSCA) that can be used for window sizes up to $2^{20}$. For both techniques, a generalised methodology is presented to parallelise the computation while respecting memory size and data bandwidth constraints.  
Compared to an NVIDIA GeForce RTX 3090 graphics processing unit (GPU) which uses a similar 7nm technology to our FPGA, for the same accuracy, our FAM/SSCA implementations achieve speedups of 4.43x/1.90x and a 30.5x/24.5x improvement in energy efficiency.
\end{abstract}


\section{Introduction}

A time series is said to be {\it cyclostationary} if its probability distribution varies periodically with time. Cyclostationary time series analyses are suitable for a wide range of periodic phenomena in signal processing, including characterisation of modulation types; noise analysis of periodic time-variant linear systems; synchronisation problems; parameter and waveform estimation; channel identification and equalisation; signal detection and classification; \gls{ar} and \gls{arma} modelling and prediction; and source separation~\cite{gardner2006cyclostationarity, 5067400, 9852206}.
Cyclostationary analysis often involves estimating the \gls{scd}, which is the idealised temporal cross-correlation between all pairs of narrowband spectral components.  

Although the \gls{scd} reveals extensive information about cyclostationary processes, the high computational requirements of the method poses problems for real-time applications. 
Even though the \gls{fft} significantly improves computational efficiency in the two most widely used \gls{scd} estimators, namely the \gls{fam} and \gls{ssca} techniques, practitioners still seek enhanced performance to detect signals buried deep within noise~\cite{Gardner91, gardner1986spectral}.
Consequently, there has been significant interest in developing high-performance implementations of the \gls{scd} method to detect and classify cyclostationary signals using \glspl{cpu}, \glspl{gpu}, and \glspl{fpga} technologies.

Traditional implementations of \gls{scd} algorithms on \glspl{cpu} and \glspl{gpu} are constrained by the fixed architectures of these platforms and are often not energy efficient. In contrast, \glspl{fpga} offer a flexible architecture that can be customised for specific applications. Moreover, they provide the possibility of integrating cyclostationary analysis with a software-defined radio and/or other signal processing and machine learning functionality. They also enable custom data paths that enhance parallelism and improve computational speed.

\begin{figure*}[t]
  \centering
  \begin{subfigure}[b]{0.48\textwidth}
    \centering
    \subcaptionbox{A real signal $x(n)$ with a sample period of $T_s$.}{
      \includegraphics[width=0.7\linewidth]{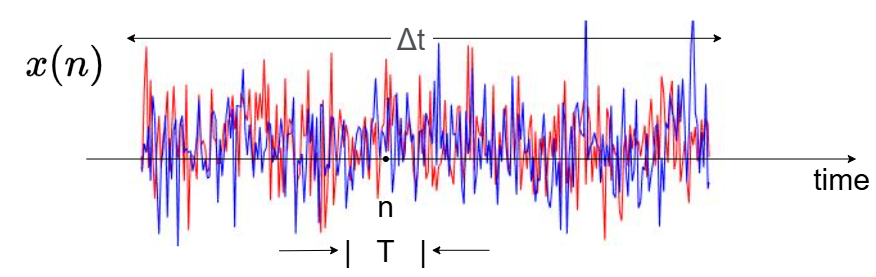}
    }
    \quad
    \subcaptionbox{Complex demodulates of signal $x(n)$.}{
      \includegraphics[width=0.7\linewidth]{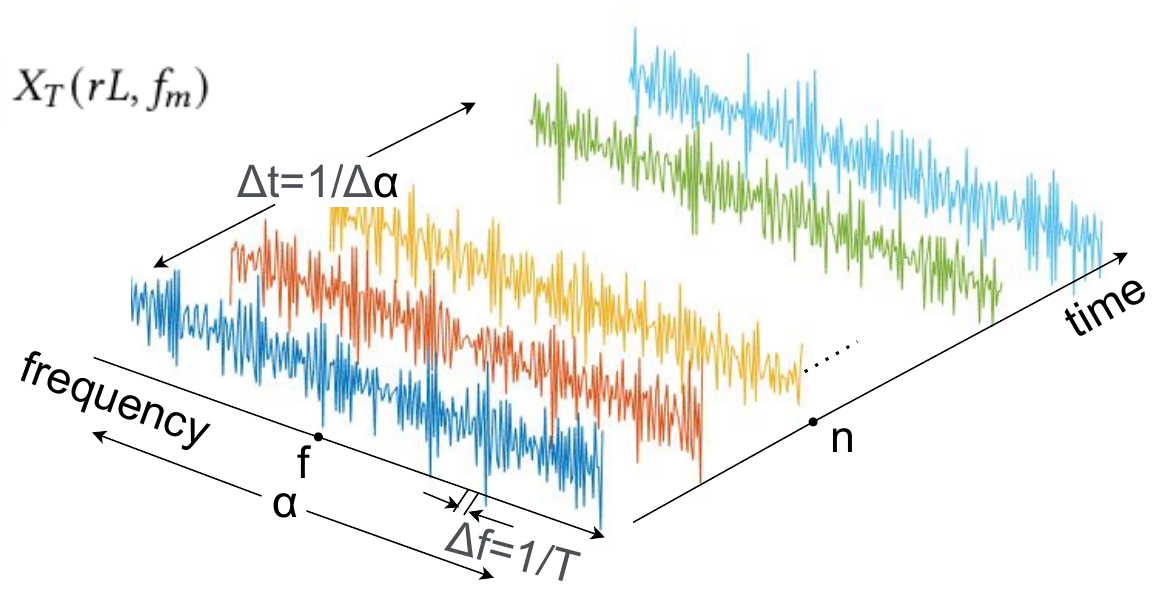}
    }
  \end{subfigure}\quad
  \begin{subfigure}[b]{0.48\textwidth}
    \centering
    \subcaptionbox{The SCD function of signal $x(n)$ with alpha profile.}{\includegraphics[width=\linewidth]{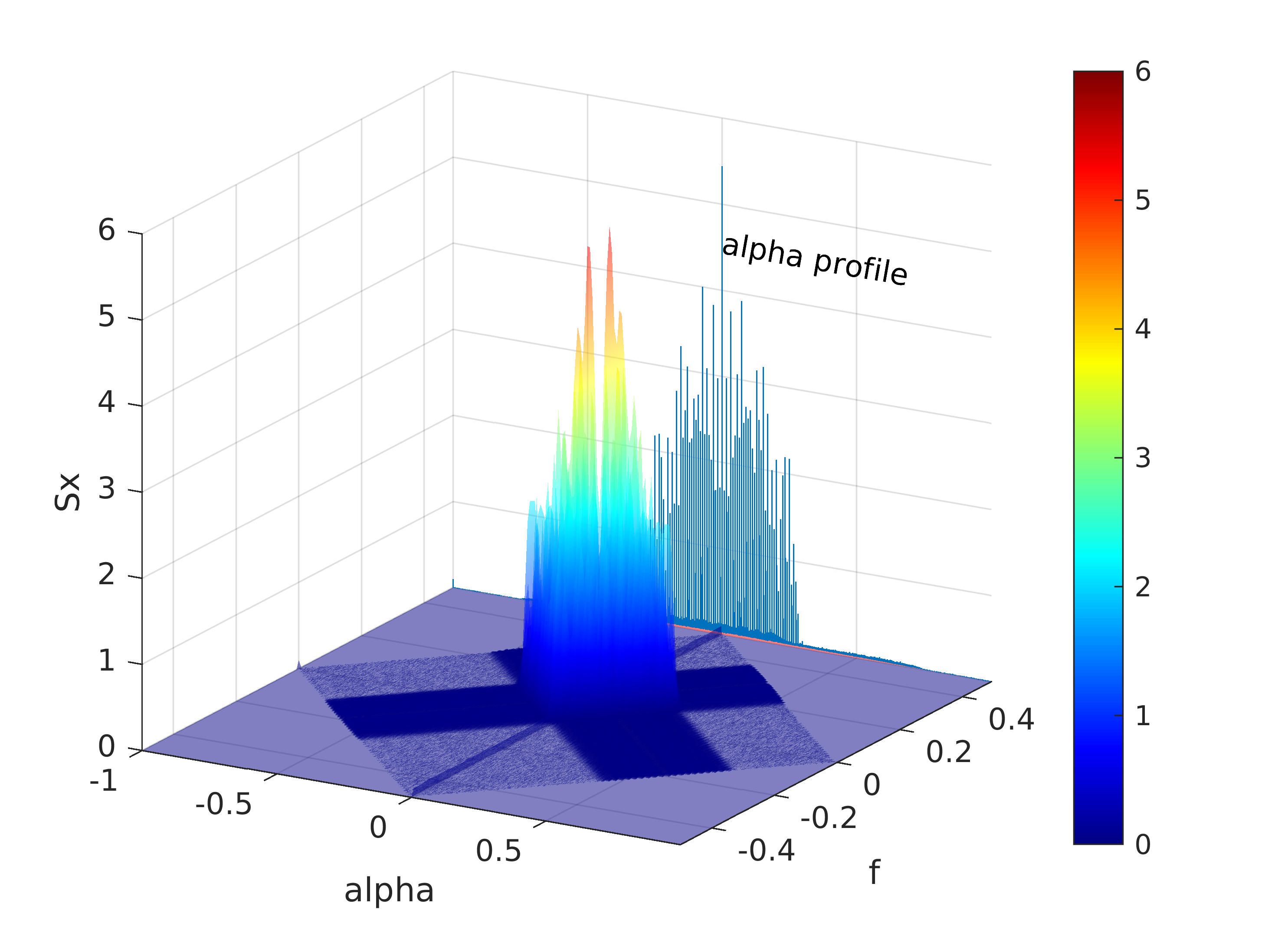}}
  \end{subfigure}
  \caption{The SCD function of \gls{dsss} \gls{bpsk} signal. }
  \vspace{-2ex}
  \label{fig:composite}
\end{figure*}
The AMD/Xilinx Versal ACAP architecture (Versal architecture), described in \cite{amd2024am009} merges general-purpose \glspl{cpu}, \gls{pl}, and \gls{aie} processors optimised for AI and machine learning optimisation. With 400 \gls{aie} processors executing at a maximum 1.25~GHz, capable of delivering 8 MACs/cycles for 32-bit floating-point data, it has peak performance of 8~\gls{tflops}~\cite{amd2024xmp452}. The \gls{scd} estimators in this work were designed to later be integrated with an \gls{sdr} front-end and \gls{ml} back-end to perform \gls{rfml}. 
The novel contributions of this paper are:
\begin{itemize}
    \item A design methodology reported for high performance  \gls{scd} estimation using the  \gls{fam} and \gls{ssca} techniques on Versal platforms.
    \item The first reported \gls{fam} implementation that only uses \glspl{aie}. Although performance is one-quarter of Ref.~\cite{10.1145/3567429}, it only requires 35\% of the \glspl{aie} available and zero \gls{pl} resources enabling future \gls{ml} integration in \gls{rfml} applications.  
    \item An SSCA implementation employing a decomposed \gls{fft} and \gls{pl} transpose unit to handle window sizes of the order of 1M samples. To the best of our knowledge, this is the first \gls{fpga}-accelerated \gls{ssca} implementation.
\end{itemize}

The remainder of this paper is structured as follows. Section~\ref{se:Background} provides an overview of the \gls{scd} algorithms and the Versal architecture. In Section~\ref{se:Method}, we detail the implementation based on \gls{aie} of two \gls{scd} algorithms on the AMD/Xilinx VCK5000 Versal platform (VCK5000). Section~\ref{se:Result} presents the experimental results, followed by the conclusions in Section~\ref{se:Conclusion}.

\section{Background}
\label{se:Background}
\subsection{Spectral Correlation Density}
The description of the SCD function below follows that of Roberts~et.~al.~\cite{roberts1991computationally} and Brown~et.~al~\cite{brown1993digital}. The discrete-time {\it complex demodulate} of a continuous time, complex-valued signal $x(t)$ at frequency $f$ is
\begin{equation}
\label{eq:II.A-complex demodulate}
    X_T(n,f)=\sum_{r=-N/2}^{N/2} a(r)x(n-r)e^{-i2\pi f(n-r)T_s}
\end{equation}
where $a(r)$ is a length $T=NT_s$ second windowing function, $T_s$ is the sampling period and $N$ is the number of samples. 
Complex demodulates are low pass sequences with bandwidths $\Delta f \approx 1/T$. 
For inputs $x(n)$ and $y(n)$ of length $N$ samples, we correlate demodulates $X_T(n,f_1)$ and $Y_T(n,f_2)$ separated by $\alpha_0$ ($f_1=f_0+\alpha_0/2$,  $f_2=f_0-\alpha_0/2$) over the time window $\Delta t=NT_s$ using a complex multiplier followed by a \gls{lpf} with bandwidth approximately $1/\Delta t$. Thus the \gls{scd} function is given by
\begin{equation}
\label{eq:II.A-SCD function}
    {S}_{xy_T}^{\alpha_0}(n,f_0)_{\Delta t} = \sum_{r} X_T(r,f_1)Y^*_T(r,f_2)g(n-r)
\end{equation}
where the $^*$ operator is a complex conjugate and $g(n)$ is a length $\Delta t=NT_s$ windowing function. For the special case of auto-correlation studied in this paper, $y(n)$ is a time-delayed value of $x(n)$, i.e., $y(n) = x(n+d)$ where $d$ is the delay. 
\begin{figure}[t]
    \centering
    \includegraphics[width=\linewidth]{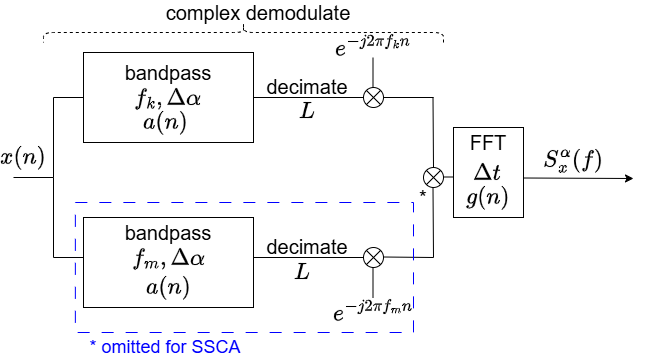}
    \caption{Dataflow for \gls{fam} and \gls{ssca} (for \gls{ssca} $L = 1$ and the dashed block is omitted).}
    \label{fig:fam_ssca_method}
\end{figure}
\begin{figure*}[t]
    \centering
    \includegraphics[width=\linewidth]{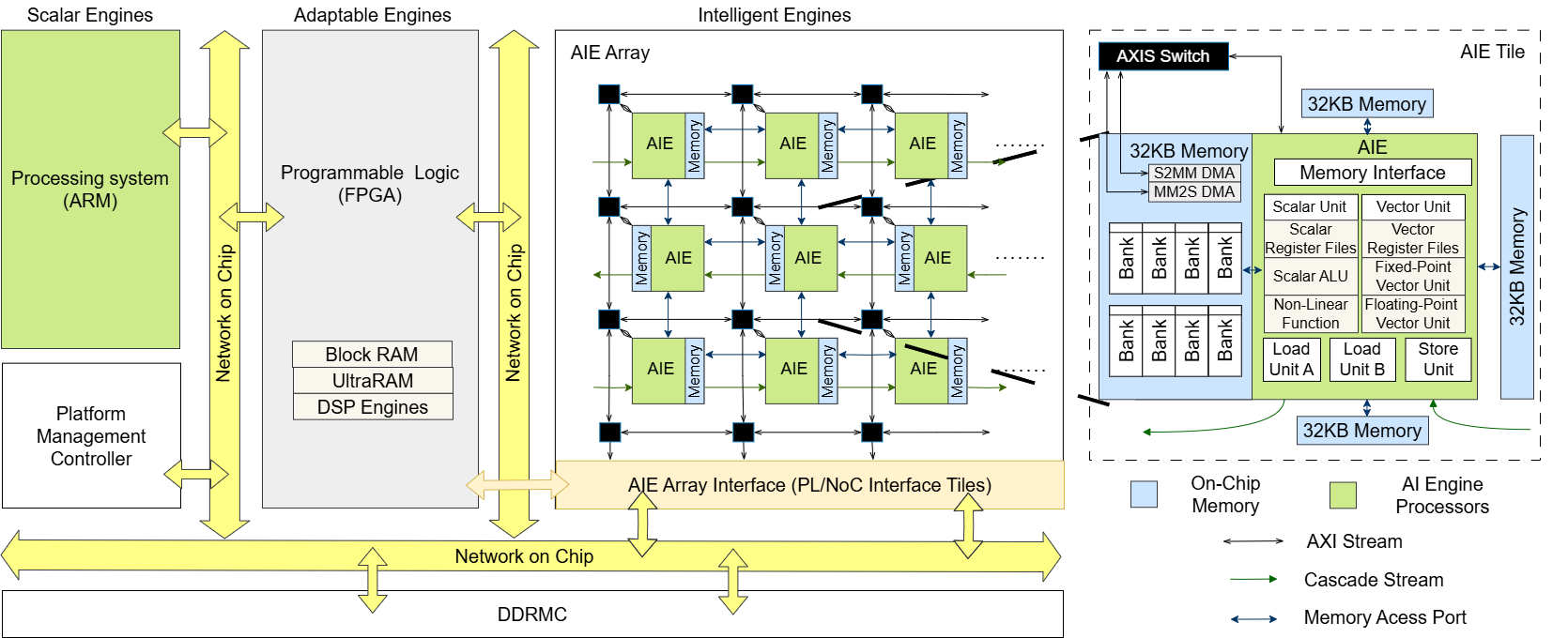}
    \caption{AMD/Xilinx Versal ACAP Architecture.}
    \label{fig:Chap4_Versal_architecture}
\end{figure*}
\subsection{FAM Technique}
\label{se:estscd}
The direct application of Eq.~\eqref{eq:II.A-SCD function} is computationally inefficient. Decimation and the \gls{fft} can be used to reduce the computational complexity~\cite{roberts1991computationally}.  Fig.~\ref{fig:fam_ssca_method} illustrates the signal flow for the \gls{fam} method, where the first task for both methods is to compute the complex demodulates, $X_T$ and $Y_T$ (in Eq.~\eqref{eq:II.A-SCD function}). We summarise the computations in this subsection, and refer readers to references~\cite{brown1993digital,gardner1994cyclostationarity} for a detailed derivation, with implementation guidance in~\cite{10.1145/3546181}.

\subsubsection{Complex Demodulate}
For Eq.~\eqref{eq:II.A-complex demodulate}, the input sequence is from $N$ to $P=N/L$ via decimation using an $L$ sample stride for the channeliser, with $L = N_P/4$~\cite{brown1993digital}. The Eq.~\eqref{eq:II.A-complex demodulate} can be rewritten as
\begin{equation}
\label{eq:II.B-complex_demodulate_final}
\begin{split}
    &X_T(pL,f_m)\\
    & =\underbrace{[\sum_{k=0}^{N_P-1} \underbrace{a(d-k)x(pL-d+k)}_{x(n)\text{ windowed by }a(n)}e^{-i2\pi mk/N_P}]}_{N_P\text{ point-FFT}}
    \underbrace{ e^{-i2\pi mpL/N_P} }_{\text{Down Conversion}},
\end{split}
\end{equation}
where $p = \{0,1...,P-1\}$, $d=N_P/2-1, r=d-k, f_m = mf_s/N_P$, $f_s = 1/T_s$, and $-N_P/2 < m < N_P/2$~\cite{gardner1994cyclostationarity}.
Thus, in Eq.~\eqref{eq:II.B-complex_demodulate_final}, the input is windowed via $a(n)$, then passed through a $N_P$-Point FFT. A phase shift is introduced to compensate for the down conversion from $N$ to $N_P$ samples.

\subsubsection{FAM method}
Taking Eq.~\eqref{eq:II.A-SCD function}, and substituting $X_T=Y_T$ to compute at the frequency $f_{kl}= (f_k+f_l)/2$ in $P$ segments (Eq.~\eqref{eq:II.B-complex_demodulate_final}),  Eq.~\eqref{eq:II.A-SCD function} becomes
\begin{equation}
\label{eq:II.B-SCD_function_auto_P_segments}
    {S}_{xy_T}^{\alpha_0}(pL,f_{kl})_{\Delta t} = \sum_{r} X_T(rL,f_k)X^*_T(rL,f_l)g_d(p-r)
\end{equation}
where $p=\{0,1,...,P-1\}$ and $g_d(r) = g(rL)$. Now the cycle frequency parameter has been redefined to $\alpha_0 = f_l-f_k+\epsilon$, as the $\epsilon=\Delta f$ is the introduced frequency shift. 


Introducing $\epsilon = q\Delta \alpha$ ($\Delta \alpha = f_s/P$ and $q=\frac{\Delta f}{\Delta \alpha}$) to Eq.~\eqref{eq:II.B-SCD_function_auto_P_segments} and
 substituting $a_{kl}=f_k-f_l$, $f_0=f_{kl}=(f_k+f_l)/2$, and $ \alpha_0 = a_{kl}+q\Delta \alpha $~\cite{gardner1994cyclostationarity}, the following is obtained
\begin{equation}
\label{eq:II.B-SCD_function_FAM}
\begin{split}
      &{S}_{x}^{a_{kl}+\Delta\alpha}(pL,f_{kl})_{\Delta t}\\
      &= \underbrace{\sum_{r}\underbrace{X_T(rL,f_k)X^*_T(rL,f_l)}_{\text{Conjugate\ Multiplication}}g_d(p-r)e^{-i2\pi rq/P}}_{\text{P-point FFT}}.
\end{split}
\end{equation}

\subsection{SSCA Technique}
Instead of multiplying two complex demodulates, the \gls{ssca} directly multiplies complex demodulates with the original signal~\cite{Brown1987phdthesis}. In Fig.~\ref{fig:fam_ssca_method}, the difference between \gls{ssca} and \gls{fam} is that instead of multiplying the complex demodulate $X_T(n,f_k)$ with $Y_T^*(n,f_k)$, it is multiplied by $y^*(n)$ to produce the \gls{cdp}, $X_g$, for $k \in [-N_P/2,N_P/2-1]$. To ensure consistency in the sampling rate of the two terms, the channeliser decimation factor $L$ is set to 1.
\begin{equation}
\label{eq:Xg}
    X_g(n+m, k) = X_T(n+m,f_k)x^*(n+m)g(m)
\end{equation}
where $g(m)$ is a length $\Delta t=NT_s$ windowing function, and $m\in [-N/2,N/2-1]$. The centre frequencies of $X_T$ are set to $f_k = k(f_s/N_P)$.

Finally, the $N$-point FFT of each of the $N_P$ \gls{cdp} values is computed resulting in the \gls{scd} estimate 
\begin{equation}
\label{eq:II.B-SSCA_auto}
    {S}_{X}^{f_k+q\Delta\alpha}(\frac{f_k}{2}-q\frac{\Delta\alpha}{2})_{\Delta t} = \underbrace{\sum_{m=-N/2}^{N/2-1} X_g(n+m,k) e^{-i2\pi q m /N}}_{\text{N-point FFT}}
\end{equation}
where cycle frequency $\alpha = f_k+q\Delta \alpha$, $\Delta \alpha = f_s/N$, $q \in [-N/2,N/2-1]$, and $f = (f_k-q\Delta \alpha)/2$~\cite{defence1994implementation,gardner1994cyclostationarity}. In the implementation, both $f$ and $\alpha$ are normalised based on $f_s=1$, which maps $S_X^{\alpha}(f)$ to $f \in [-0.5, 0.5]$ and $\alpha \in [-1,1]$.

\subsection{Versal Architecture}

\begin{figure*}[!t]
    \centering
    \includegraphics[width=0.95\linewidth]{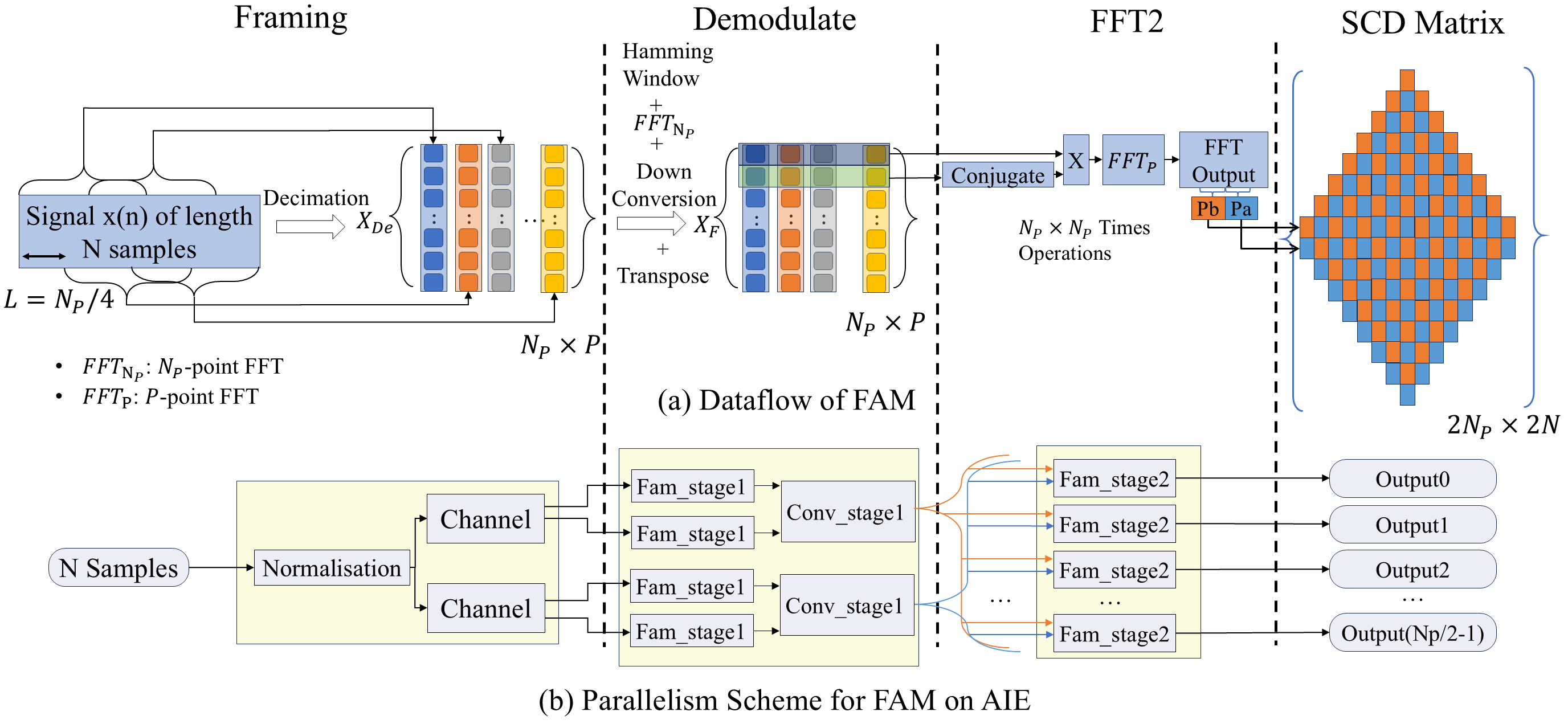}
    \caption{Dataflow and parallelism scheme for FAM on VCK5000.}
\vspace{-1ex}
    \label{fig:Chap4_fam_arch}
\end{figure*}

The AMD/Xilinx VCK5000 development card is powered by AMD's 7nm Versal\texttrademark{} Versal XCVC1902-2MSEVSVA2197 Adaptive SoC (VC1902), and \gls{aie} development is facilitated by the Vitis software platform~\cite{netVCK5000}.

Fig.~\ref{fig:Chap4_Versal_architecture} presents the overall Versal architecture, emphasising the \gls{aie} tile on the right~\cite{amd2024am009}. 
The \gls{pl} can be customised to meet specific application requirements, with \gls{dsp} capabilities integrated for enhanced functionality.
Additionally, the board incorporates an ARM processor for general-purpose processing tasks. The \gls{aie} array supports C/C++ programmability. The \gls{pl} design can be made using RTL or C/C++ via \gls{hls}.~\cite{amdUG1076}

These three components, the \gls{aie} array, ARM processor, and \gls{pl}, are integral parts of the heterogeneous SoC. They operate independently and communicate through a \gls{noc} to other peripherals such as PCIe and DRAM controllers. 
The VCK5000 additionally features four DDR4 off-chip memory modules, each providing a peak bandwidth of 25.6~GB/s~\cite{amd2024am009}.

\section{Method}
\label{se:Method}
The \gls{fam} method is limited by its nonuniform frequency resolution, making it suitable for analysing small-window signals where resolution is less critical~\cite{roberts1991computationally, ANTONI2017248}. In contrast, the \gls{ssca} method provides uniform frequency resolution, making it appropriate for large-window signals analysis, with higher memory requirements~\cite{roberts1991computationally, brown1993digital}. 
In this section, we present an \gls{aie}-based small-window \gls{fam} implementation optimised for processing speed (Sec.~\ref{se:fam_imp}), and a large-window \gls{ssca} implementation for applications that require accurate cycle frequency estimates (Sec.~\ref{se:ssca_imp}).
\vspace{-1ex}
\subsection{FAM Implementation}
\label{se:fam_imp}
Implementing the \gls{fam} entirely within \gls{aie} tiles allows for faster internal data transfers, avoiding a potential bottleneck between the \gls{pl} and \gls{aie}. Thus, we focus on efficient utilisation of the limited \gls{aie} tile memory that is available. Key hardware constraints include:

\begin{itemize}

\item \textbf{\gls{aie} Tile Memory Constraints.}%
\ While an \gls{aie} tile can access data memory in four directions, a maximum of 2 input AXI4-Stream interfaces are supported, limiting the number of input buffers to two, i.e. the total input buffer capacity is
\(2 \times 32\,\text{KB} = 64\,\text{KB}\).
Since each input buffer is organised in ping–pong A/B halves to overlap writes with computations, the actual size is halved again to 32\,\text{KB}.
 Therefore, the usable size for each individual input buffer on an \gls{aie} Tile is 16 KB.

\item \textbf{\gls{aie} array to \gls{pl} Interface Constraints.} The VC1902 provides 39 columns to interface between the \gls{aie} array and PL, with each column having 6 streams (totaling 234 stream interfaces)~\cite{amd2024am009}.
\end{itemize}

Fig.~\ref{fig:Chap4_fam_arch} illustrates the \gls{fam} process from input samples to the final  SCD matrix, covering three stages: \textbf{Framing}, \textbf{Demodulate}, and \textbf{FFT2}. Below, we analyse the relationships between parameters and optimisation details for the typical settings used in previous work~\cite{10.1145/3546181, 10.1145/3567429} ($N=2,048$, $N_P=256$) to facilitate an effective comparison.

\subsubsection{Methodology}
\label{ch:ch3_fam_method}
The small-window \gls{fam} architecture supports $N_P$ values from $2^4$ to $2^8$, and input signal length $N$ from $2^7$ to $2^{12}$. We define
the total number of \gls{aie} tiles 
required to compute $X_T$ in Eq.~\ref{eq:II.B-complex_demodulate_final} as:
\vspace{-0.5ex}
\begin{equation}
\mathbb{A}_{FAM} = \mathbb{A}_{Framing} + \mathbb{A}_{Demodulate} + \mathbb{A}_{FFT2},
\label{eq:AIENO_FAM}
\vspace{-0.5ex}
\end{equation}
where $\mathbb{A}_{Framing}$, $\mathbb{A}_{Demodulate}$ and $\mathbb{A}_{FFT2}$ are the number of tiles required for the Framing, Demodulate and FFT2 stages respectively.
The intermediate matrix $X_F$ between the demodulate stage and the FFT2 stage has a size of $N_P \times P$ complex values. Usually we set $L = N_P/4$~\cite{brown1993digital}, so it follows that $P = N/L = 4N/N_P$. Therefore, the total size of the intermediate matrix $X_F$ becomes $N_P P = 4N$ complex numbers. Each input buffer (up to 16 KB) stores $F=2,048$ complex floats, needing $\lceil4N/F
\rceil$ buffers and \texttt{Fam\_stage1} kernels in the Demodulate stage, plus $\lceil4N/(2F)
\rceil$ \texttt{Conv\_stage1} kernels to aggregate
data from the buffers, where $\lceil \cdot \rceil$ denotes the ceiling function. 

The Framing stage employs one \texttt{norm} kernel for normalisation and $\lceil4N/(2F)
\rceil$ \texttt{Channel} kernels to distribute data to Demodulate kernels.
The FFT2 stage involves conjugate multiplications and \gls{fft} computations. With a maximum of 234 \gls{aie}-to-PL streams, if $N_P>234$ (e.g., 256), 128 kernels perform $N_P /128 \times N_P$ operations each, outputting results to 128 streams.
Thus, the total tile count is:
\vspace{-0.5ex}
\begin{equation}
\label{eq:AIENUMFAM}
\mathbb{A}_{FAM} = 1 + \left\lceil\frac{4N}{2F}\right\rceil +\left(\left\lceil\frac{4N}{2F}\right\rceil+\left\lceil\frac{4N}{F}\right\rceil\right)+\min(N_P,128)
\end{equation}

Next, we will introduce the specific implementation of \gls{fam} on \gls{aie} array when $N = 2,048, N_P = 256$, and $P = 32$.

\subsubsection{Framing Stage}
Before decimation, the \texttt{norm} kernel normalises input \( x(t) \). According to Eq.~\eqref{eq:II.B-complex_demodulate_final}, the signal of length \( N \) is partitioned into \( P \) blocks, each containing \( N_P \) elements with offset \( L = N_P/4 \). The decimated matrix \( X_{\text{De}} \in \mathbb{R}^{N_P \times P} \) is defined as:
\[X_{\text{De}}[n, p] = x[pL + n], \quad 0 \le n < N_P,\; 0 \le p < P.\]

The matrix \(X_{\text{De}}\) is column-wise partitioned into four equal submatrices \(X_{\text{De}}^{(i)} \in \mathbb{C}^{N_P \times P/4}\) for parallel processing:
\[
X_{\text{De}} = \left[ X_{\text{De}}^{(0)} | X_{\text{De}}^{(1)} | X_{\text{De}}^{(2)} | X_{\text{De}}^{(3)} \right],
\]
each mapped to independent kernels. The \texttt{channel} kernel routes data to Demodulate stage tiles.

\subsubsection{Demodulate Stage}
Based on Eq.~\eqref{eq:II.B-complex_demodulate_final}, each column of the matrix from the Framing stage undergoes windowing, down conversion, and \(N_P\)-point \gls{fft}. Considering memory constraints (32KB per tile), this stage uses four \texttt{Fam\_stage1} kernels, each processing \(P/4\) frames, and two \texttt{Conv\_stage1} kernels to broadcast data for FFT2.

Each kernel \( i \) processes frames in index set \(\mathcal{J}_i\):
\[
\mathcal{J}_i = \left\{ j \,|\, i\frac{P}{4} \le j < (i+1)\frac{P}{4} \right\}, \quad i=0,1,2,3.
\]
Kernel outputs \(X_T^{(i)}(f_m)\) correspond to these frame subsets:
\vspace{-0.5ex}
\[
X_T^{(i)}(f_m) = [\, X_T(jL, f_m) ]_{j\in \mathcal{J}_i}, \quad i=0,1,2,3,
\]
forming the combined output:
\vspace{-0.5ex}
\[
X_T(f_m) = [\,X_T^{(0)}(f_m) | X_T^{(1)}(f_m) | X_T^{(2)}(f_m) | X_T^{(3)}(f_m)\,].
\]
In our design, the output from the Demodulate stage forms an \(N_P \times P\) matrix, distributed across four kernels, each handling \(P/4\) columns. Each \texttt{FAM\_Stage1} kernel stores its column data into a buffer with a \(P/4\)-element stride. Thus, every \(P/4\)-element block in memory corresponds to \(P/4\) points from one matrix row. Two \texttt{Conv} kernels then reorganise these blocks into four contiguous groups of \(P/4\) points. The \texttt{FAM\_Stage2} kernel receives the complete \(P\)-point input through two ports.

%

\begin{algorithm}[ht]
\small 
\caption{\gls{aie}-based \gls{fam} pipeline pseudocode.}
\label{alg:FAM_pseudocode}
\SetKwComment{Comment}{$\triangleright$\ }{}
\SetAlgoNlRelativeSize{0}
\SetKwProg{Fn}{Function}{:}{}
\SetKwFunction{framing}{Framing}
\SetKwFunction{demodulate}{Demodulate}
\SetKwFunction{fft}{FFT2}
\SetKwFunction{normalise}{Normalise}
\SetKwFunction{chanone}{Channel}
\SetKwFunction{chantwo}{Channel2}
\SetKwFunction{famone}{FamStage1}
\SetKwFunction{convone}{ConvStage1}
\SetKwFunction{famtwo}{FamStage2}

\Fn{\framing{$data\_in$}}{
  $data\_norm \gets Normalize(data\_in) $\;
  $X \gets \text{zeros}(N_P, P)$\;
  \For{$k = 0$ \KwTo $P-1$}{
    $X(:,k+1) \gets data\_norm[k \cdot L : k \cdot L + N_P - 1]$\;
  }
  \KwRet $X$\;
}
\Fn{\demodulate{$X$}}{
  $Y \gets \text{zeros}(N_P, P)$\;
  \For{$k = 0$ \KwTo $P-1$}{
    $data\_win \gets \text{Window}(chebwin[N_P], X(:,k))$\;
    $data\_fft \gets N_P\text{-Point FFT}(data\_win)$\;
    $Y(:,k) \gets \text{Down\_Conversion}(data\_fft,\, k)$\;
  }
  \KwRet $Y$\;
}
\Fn{\fft{$Y$}}{
  \For{$m = 0$ \KwTo $N_P - 1$}{
    \For{$n = 0$ \KwTo $N_P - 1$}{
      $z \gets Y(:,m) \cdot \text{conj}(Y(:,n))$\;
      $z_{fft} \gets P\text{-point FFT}(z)$\;
      $z_{abs} \gets |z_{fft}|^2$\;
      write\_to\_output($z_{abs}[\tfrac{P}{2}.. \tfrac{3P}{4}-1]$, $z_{abs}[\tfrac{P}{4}.. \tfrac{P}{2}-1]$)\;
    }
  }
}
\end{algorithm}

\subsubsection{FFT2 Stage}
For \(N_P=256\), 128 \texttt{\gls{fam}\_stage2} kernels handle two frequency channels each:
\vspace{-0.5ex}
\[
f_{k_1}=2i,\quad f_{k_2}=2i+1,\quad i=0,\dots,127.
\]
with the corresponding data \(\mathbf{X}_T(rL, f_{k_1})\) and \(\mathbf{X}_T(rL, f_{k_2})\) pre-loaded from \texttt{Conv\_stage1} kernels. In FFT2 stage, the complex-conjugate data \(\mathbf{X}_T^*(rL,f_l)\) is then broadcast \emph{twice} to each kernel:

\begin{enumerate}
  \item \textbf{Broadcast 1:} Multiply \(\mathbf{X}_T(rL,f_{k_1})\) with \(\mathbf{X}_T^*(rL,f_l)\).
  \item \textbf{Broadcast 2:} Multiply \(\mathbf{X}_T(rL,f_{k_2})\) with \(\mathbf{X}_T^*(rL,f_l)\).
\end{enumerate}

Mathematically, this expands the original single-pair operation \(\bigl\{(k,l)\bigr\}\) into two passes for the pair \(\{k_1,k_2\}\). For each \(\texttt{FAM\_stage2}\) kernel, Eq.~\eqref{eq:II.B-SCD_function_FAM} becomes:
\vspace{-0.5ex}
\[
S_x^{(k_j,l)}(pL)=\sum_{r=0}^{P-1}
X_T(rL,f_{k_j})X_T^*(rL,f_l)\,g_d(p-r)e^{-i2\pi rq/P}.
\]
By performing two broadcast passes of \(\mathbf{X}_T^*(rL,f_l)\), each kernel can process the conjugate multiplications for its two assigned frequency channels \(\{f_{k_1}, f_{k_2}\}\) without exceeding the available I/O resources.

\subsubsection{Implementation}
We implement the \gls{fam} algorithm on VCK5000 platform. \gls{pl} fetches 2,048 complex samples (64-bit, 8\,KB) from DDR and streams them as two 32-bit AXI streams to \gls{aie} tiles. The entire algorithm is presented in Alg.~\ref{alg:FAM_pseudocode}.

\vspace{-0.5ex}
\subsection{SSCA Implementation}
\label{se:ssca_imp}
For very large \gls{ssca} windows (e.g., $N=2^{20}$), intermediate matrices $X_g$ must be stored on- or off-chip, and the \gls{aie} array primarily addresses computational complexity and store temporary variables. The key challenge on the VCK5000 is implementing an $N$-point \gls{fft} while managing memory bandwidth constraints. 

\subsubsection{SSCA\_2DFFT}
The \gls{2dssca} is a novel approach that implements \gls{ssca} with a \gls{2dfft}, aligning the \gls{cdp} computation order with the input sequence expected by the \gls{2dfft}. This decomposition enables a more efficient mapping onto the \gls{aie} array, reducing intermediate matrix size and conserving memory bandwidth.

I.~J.~Good provided~\cite{10.1111/j.2517-6161.1958.tb00300.x} the theoretical foundation for computing the Fourier transform of one-dimensional signals via multi-dimensional Fourier transforms~\cite{alma991004238809705106}. The $N$-point \gls{dft} can be computed using an $M_1M_2$-point matrix \gls{dft} as:
\begin{equation}
\vspace{-1.5ex}
\label{eq:Chap4_2DFFT}
\begin{split}
    &\hat{x}(m_1',m_2')\\ 
    &=\underbrace{\sum^{M_2-1}_{m_2=0}\underbrace{\{\sum^{M_1-1}_{m_1=0}x(m_1,m_2)e^{-j2\pi\frac{m_1m_1'}{M_1}}\}}_{M_1-\text{point DFT on $m_2$th column}}\underbrace{e^{-j2\pi \frac{m_2m_1'}{M_2M_1}}}_{\text{rotate factor}}e^{-j2\pi \frac{m_2m_2'}{M_2}}}_{M_2-\text{point DFT on $m_1$th row}},
\end{split}
\end{equation}
where $M_1M_2 = N$, $0\leq m_1,\,m_1'<M_1$, and $0\leq m_2,\,m_2'< M_2$.

\begin{figure}[t]
    \centering
    \includegraphics[width=\linewidth]{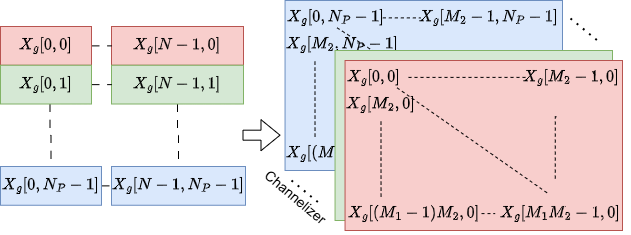}
    \caption{Reshaping $ X_g $ from $[N \times N_P]$ to $[M_1 \times M_2 \times N_P]$, where $N = M_1 M_2$.}
    \vspace{-2ex}
    \label{fig:Chap4_2DFFT_Xg}
\end{figure}
Fig.~\ref{fig:Chap4_2DFFT_Xg} shows how \(X_g\) is reshaped in \gls{2dssca} to serve as input for the decomposed \gls{fft} stages, by mapping the 1D index to 2D coordinates for data reorganisation:
\vspace{-0.5ex}
\begin{equation}
    X_g^{2D}(m_1, m_2, k )
        =X_g(m_1M_2+m_2, k),
\end{equation}
where $0\leq m_1<M_1$, $0\leq m_2< M_2$ and $0\leq k<N_P$.
The \(N_P\) channelizer computations run in parallel within \(X_g\) and can overlap with the \gls{2dfft}, so the \(N_P\) \glspl{2dfft} can be treated collectively, generating \(X_g\) only when needed.

The \gls{2dssca} operates in two stages. \textbf{Stage 1} performs $M_1$-point \glspl{fft} on each of the $M_1$ rows of $X_g$, across all $M_2$ columns for $N_P$ channelizers. During this stage, $X_g$ is provided as required, and the outputs are then multiplied by a rotation factor and stored. \textbf{Stage 2} uses the results from Stage 1 to perform the $M_2$-point \gls{fft} for each row and channelizer.
To achieve this, Eq.~\eqref{eq:II.B-complex_demodulate_final} becomes:
\vspace{-0.5ex}
\begin{equation}
    \label{eq:Chap4_XT_2}
    \begin{split}
        &X_T^{2D}(m_1, m_2, f) = X_T(m_1 M_2 + m_2,f)\\
        &= \sum_{r=-N_P/2}^{N_P/2-1} a(r) x(\eta+ r) e^{-j 2 \pi f r T_s}*e^{-j 2 \pi f \eta T_s}
    \end{split}
\end{equation}
where $\eta$ is replaced by $\eta = m_1 M_2 + m_2$. 
If $M_2$ is divisible by $N_P$, the down conversion term becomes $e^{-i2\pi fm_2T_s}$ for all $m_1$. 
After multiplying by $x^*(\eta)$, the \gls{cdp} is expressed as:
\begin{equation}
    \label{Chap4_Xg_2}
        X_g^{2D}(m_1, m_2, k )
        = X_T^{2D}(m_1, m_2, f_k) x^*(\eta + m) g(m).
\end{equation}
The final \gls{2dssca} is then obtained by computing the \gls{2dfft} of the $N_P$ \gls{cdp} values:
\begin{equation}
\label{eq:Chap4_SSCA_2DFFT}
    {S}_{x}^{2D}(m_1', m_2', k)_{\Delta t} = \textbf{Stage 2}(\textbf{Stage 1($X_g^{2D}(m_1, m_2, k)$)}),
\end{equation}
in which the \textbf{Stage 1} is 
\begin{equation}
\begin{split}
        &{S}_{x.s1}^{2D}(m_1', m_2, k)_{\Delta t}\\& = \underbrace{\sum^{M_1-1}_{m_1=0} X_g^{2D}(m_1, m_2, k) e^{-j 2 \pi \frac{m_1 m_1'}{M_1}}}_{\text{$M_1$-point FFT}} \underbrace{e^{-j 2 \pi \frac{m_2 m_1'}{M_2 M_1}}}_{\text{rotate factor}}
\end{split}
\end{equation}
and the \textbf{Stage 2} is
\begin{equation}
    {S}_{x}^{2D}(m_1', m_2', k)_{\Delta t} = \underbrace{\sum^{M_2-1}_{m_2=0}{S}_{x.s1}^{2D}(m_1', m_2, k)_{\Delta t}e^{-j 2 \pi \frac{m_2 m_2'}{M_2}}}_{\text{$M_2$-point FFT}}.
\end{equation}

Finally, $S_x^{2D}(m_1, m_2, k)$ is mapped to $S^\alpha_x(f)$ using:
\[
f = \frac{k}{2 N_P} - \frac{M_1 m_2' + m_1' - N/2}{2 N}
\]
\[
\alpha = \frac{k}{N_P} + \frac{M_1 m_2' + m_1' - N/2}{N}.
\]

\begin{figure}[t]
    \centering
    \includegraphics[width=\linewidth]{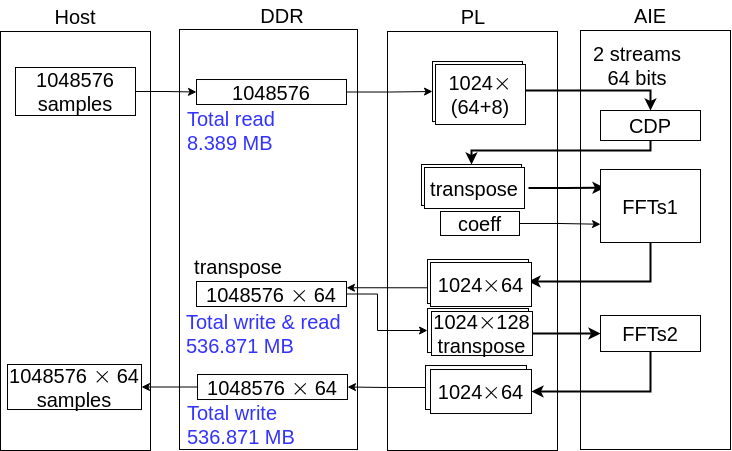}
    \caption{Dataflow of SSCA\_2DFFT on Versal.}
    \vspace{-2ex}
    \label{fig:Chap4_SSCA_Versal_dataflow}
\end{figure}
\subsubsection{Methodology}
The VCK5000 platform offers 23.9~MB of on-chip SRAM and four 4~GB off-chip DDR. Since an intermediate matrix in single-precision complex format requires $8\times N\times N_P$ bytes, off-chip memory is needed if $N\times N_P > 2^{20}$.

Each \gls{aie} tile features eight single-port memory banks (32KB total), sufficient to perform two-step 1K (1,024-element) single-precision complex computations using a ping-pong buffer scheme. In the \gls{ssca} implementation, the number of $N_P$ channelizers ranges from $2^5$ to $2^8$, and the size of the second $N$-point \gls{fft} spans from $2^{12}$ to $2^{20}$. Because each \gls{aie} tile is efficient at processing a 1K array, we assign $\mathbb{A}_{CDP}$ \gls{aie} tiles to compute $1K/N_P$ window sets of $X_g$ (Eq.~\ref{eq:Xg}) in parallel. 
Accordingly, $\mathbb{A}_{CDP}$ is given by:
\begin{equation}
        \mathbb{A}_{CDP}= 1+\lceil\log _2(N_P)/2\rceil,
\end{equation}
with one for down conversion/conjugate multiplication, and the rest tiles for $N_P$-point \gls{fft}. The window function is merged into the first \gls{fft} stage.

For the large-point \gls{fft}, the $N$-point \gls{fft} is decomposed into $M_1$ and $M_2 = N/M_1$. Thus, the number of \gls{aie} tiles required is:
\begin{equation}
    \mathbb{A}_{2DFFT} = \lceil\log_2(M_1)/2\rceil + 1 +\lceil\log_2(M_2)/2\rceil,
\end{equation}
where $M_1,~M_2<1K$. Additionally, one extra tile is allocated for multiplication with the rotation factors. The design computes $1K/M_1$ instances of $M_1$-point \gls{fft} and  $1K/M_2$ instances of $M_2$-point \gls{fft} in parallel. 

Thus, the total \gls{aie} tiles needed for \gls{2dssca} is
\begin{equation}
     \mathbb{A}_{SSCA} = \mathbb{A}_{CDP} + \mathbb{A}_{2DFFT}.
     \label{eq:AIENO_SSCA}
\end{equation}
Replicating $\mathbb{A}_{SSCA}$ modules enables further parallelisation.

\vspace{-1ex}
\subsubsection{Implementation}
\label{se:Chap4_Design}
We developed a complete \gls{ssca} implementation with $N = 1,048,576$ (i.e., an example of one million input size), $N_P = 64$, and $M_1 = M_2 = 1,024$ to balance the number of iterations between \texttt{FFTs1} and \texttt{FFTs2}. Fig.~\ref{fig:Chap4_SSCA_Versal_dataflow} shows the overall architecture on the VCK5000 platform. The \gls{pl} supports data transfer from the \gls{ddrmc} to the \gls{aie} array, and between sections in \gls{aie} tiles. Our implementation of a single-precision, large N-point \gls{fft} uses ideas from Ref.~\cite{FFT1M} but: (1) loads the data from \gls{cdp}, (2) matches the bandwidth of the \gls{ddrmc}, and (3) minimises DDR access time by optimising row accesses. Previous works such as Ref.~\cite{8734795} used fixed-point arithmetic.

Alg.~\ref{alg:Chap4_aie_function} outlines the \gls{2dssca} implementation on the \gls{aie} array. The \gls{pl} streams input to the \texttt{CDP} stage, which performs windowing, $N_P$-point \glspl{fft}, down conversion, and conjugate multiplication. The $M_1 \times N_P$ output matrix is transposed in \gls{pl} and forwarded to \texttt{FFTs1}, where $M_1$-point \glspl{fft} and rotate factors are applied across $M_2$ iterations. Results exceeding on-chip memory capacity ($N_P \times M_1\times M_2$) are stored in DDR. Then \gls{pl} reads the data for \texttt{FFTs2} using a stride access pattern to achieve transpose input. 

\begin{algorithm}[t]
\small
\caption{{AIE}-based \gls{2dssca} pseudocode.}
\label{alg:Chap4_aie_function}
\SetKwFunction{cdp}{CDP}
\SetKwFunction{ffts}{FFTs1}
\SetKwFunction{fftss}{FFTs2}
\SetKwComment{Comment}{$\triangleright$\ }{}

\SetAlgoNlRelativeSize{0}
\SetKwProg{Fn}{Function}{:}{}
\Fn(){\cdp{$datain\_even, datain\_odd$}}{
    $data\_win,\, data\_fft,\, data\_dc,\,data\_out = zeros(M_1,1)$;\Comment*[f]{$M_1 = 16N_P$}\\
    $xc = zeros(16,1)$\;
    static int $itr = 0$\;
    
    $data\_win, xc \gets \text{Window}(chebwin[N_P], datain\_even, datain\_odd)$\;
    $data\_fft \gets \text{$\mathbf{N_P}$-dimensional FFT}(data\_win)$;\\ 
    $data\_dc \gets \text{Down\_Conversion}(data\_fft, \text{int}(itr/N_P))$;\\ 
    $data\_out = data\_dc \times xc$;\\
    $itr = (itr==(M_2*N_P))?\, 0:\, (itr+1)$\;
    \KwRet $data\_out$\;
}
\Fn(){\ffts{$datain\_even, datain\_odd, tow$}}{
$data\_fft,\,data\_out,\,rotate\_factor = zeros(M_1,1)$\;
    \If{$itr\%N_P==0$}{
    $rotate\_factor \gets Compute\_rotate\_factor(tow) $\;
    }
    $data\_fft \gets \text{$\mathbf{M_1}$-dimensional FFT}(data\_even,\,data\_odd)$;\\ 
    $data\_out = data\_fft \times rotate\_factor $\;
    $itr = (itr==(M_2*N_P))?\, 0:\, (itr+1)$\;
    \KwRet $data\_out$\;
}
\Fn(){\fftss{$datain\_even, datain\_odd, tow$}}{
$data\_out = zeros(M_2,1)$\;
    $data\_out \gets \text{$\mathbf{M_2}$-dimensional FFT}(data\_even,\,data\_odd)$;\\ 
    \KwRet $data\_out$\;
}
\end{algorithm}

The \gls{pl} manages data transfers between \gls{ddrmc} and \gls{aie} array. 
To fully utilise the bandwidth, a 512-bit data bus transfers eight 64-bit complex samples per cycle. For \texttt{CDP}, the \gls{pl} uses a ping-pong scheme using two algernative buffers $B_0$ and $B_1\in \mathbb{C}^{M_1\times (N_P+LANE)}$, where $LANE = 8$. While one buffer is filled by the \gls{pl}, the other is read by the \gls{aie}, alternating each batch. The buffer accommodates $LANE\cdot N_P$ iterations for \texttt{CDP} per load, ensuring continuous data supply to the \texttt{CDP} module with minimal stalling.

To match the input format of \texttt{FFTs1}, the \texttt{CDP} output (an $[M_1 \times N_P]$ matrix) is transposed in the \gls{pl}. A ping-pong buffer enables parallel loading and forwarding, hiding latency and maintaining continuous dataflow to \texttt{FFTs1}.

The \gls{pl} also manages ping-pong buffers for storing \texttt{FFTs1} outputs and loading inputs to \texttt{FFTs2} via DDR. While storage to DDR is sequential, loading requires transposed access. To avoid inefficient strided reads, we allocated a larger buffer to fetch blocks of size [$M_2\times CN_P$] instead of [$M_2\times N_P$], improving access efficiency by a factor of $C$.

\begin{table*}[]
\centering
\caption{Utilisation in VCK5000}
\label{tab:Chap4_utilize}
\begin{tabular}{|l|l|l|l|l|l|l|l|}
\hline
 & \multicolumn{5}{c|}{PL}& \multicolumn{2}{c|}{\gls{aie} Array}\\ \hline
                & Register         & LUT               & LUT as MEM       & BRAM           & URAM       & \gls{aie} tile  & PLIO   \\ \hline
Total resources & 1,739,432        & 860,336           & 446,367        & 933            & 463          & 400 & - \\ \hline

FAM          & 113,686 (6.61\%) & 107,601 (12.73\%) & 960 (0.22\%)   & 37 (3.97\%)    & 0 (0.00\%)    &137 (34.25\%) & 130\\ \hline
SSCA\_2DFFT & 15,475 (0.89\%)   & 11,824 (1.37\%)    & 1,575 (0.35\%)  & 349 (37.41\%)  & 192 (41.47\%) &15 (3.75\%) &13\\ \hline
\end{tabular}%
\end{table*}
\section{Results}
\label{se:Result}
We implement the \gls{fam} and \gls{ssca} designs described in Sec.~\ref{se:Method} on the VCK5000 platform, with \gls{aie} running at 1~GHz and \gls{pl} running at 312.5~MHz. We then compare their performance against a conventional implementation on an Intel(R) Xeon(R) Silver 4208 CPU and NVIDIA GeForce RTX 3090, both at 2.10~GHz under Ubuntu 22.04.4 LTS.

\subsection{Accuracy}


Accuracy was tested using a \gls{dsss} \gls{bpsk} signal with 10~dB \gls{snr}, processing gain of 31, chip rate 0.25 and sample rate normalised to 1, resulting in cycle frequencies that are multiples of the data rate (0.25/31). 
We used IEEE 754 double-precision MATLAB results as a reference for validation. When comparing the MATLAB and VCK5000 implementations, the FAM algorithm achieves an average relative error of 9.94e-5.
We compare the VCK5000-based \gls{2dssca}  with a \gls{cpu} \gls{ssca} implementation written in C++ that used the same  \gls{fft} coefficients. Under these conditions, the average relative error reduced to 1.08e-6.

\subsection{Utilisation}
Tab.~\ref{tab:Chap4_utilize} shows the utilisation of resources. The FAM design does not require URAM, as buffering is managed within \gls{aie} tiles. For the \gls{ssca}, BRAM and URAM are used, mainly for the ping-pong buffer between the \gls{ddrmc} and \gls{aie} components. 

In the \gls{fam} implementation, 3 \gls{aie} tiles are allocated for signal normalisation and decimation in the Framing stage. The Demodulate stage employs 6 \gls{aie} tiles to perform windowing, $N_P$-point FFT, and down conversion. In the final FFT2 stage, since $N_P > 128$, 128 \gls{aie} tiles are used to execute the conjugate multiplication and $P$-point FFT. As shown in Tab.~\ref{tab:Chap4_utilize}, the implementation of the FAM algorithm requires 137 \gls{aie} tiles, which is consistent with the value in Eq.~\eqref{eq:AIENUMFAM}. In this case, all computational kernels are executed on the \gls{aie} array, making the design easily portable to \gls{aie}-only platforms.

In the \gls{ssca} implementation, the \texttt{CDP} module requires $4$ \gls{aie} tiles that compute $M_1/N_P$ sets of $N_P$ data in each iteration, and this utilises a ping-pong buffering scheme to exchange data between tiles. In the \texttt{FFTs1} and \texttt{FFTs2} modules, $5$ \gls{aie} tiles are used to compute the $M_1$-point \gls{fft}, with an additional tile in \texttt{FFTs1} for the computation of rotation factors. This totals 15 \gls{aie} tiles (Tab.~\ref{tab:Chap4_utilize}), matching  Eq.~\eqref{eq:AIENO_SSCA}.
 

\subsection{Performance}

\begin{table}
\centering
\caption{Comparison with other FAM implementations }
\label{tab:Comparison}
 \begin{threeparttable}
\begin{tabular}{|l|l|l|l|}
\hline
        & \cite{10.1145/3546181} & \cite{10.1145/3567429} & Our\\ \hline
Platform     & ZCU111         & ZCU111             & VCK5000                        \\ \hline
Initiation Interval (ms)     & 0.26        & 0.164         & 0.63        \\ \hline
Throughput (MS/s) & 7.88         & 12.50       & 3.25                   \\ \hline

Computational Performance (GOPS) & 60.40         & 460       & 189                  \\ \hline

Board Power (W) & 12.50\tnote{1}       & 35       & 40                   \\ \hline

\end{tabular}%
\begin{tablenotes}
    \item[1] Chip rather than board power. 
\end{tablenotes}
 \end{threeparttable}
\end{table}

\begin{table}[t]
\centering
\caption{Execution time and speedup vs CPU and GPU}
\label{tab:Chap4_speedup1}
\resizebox{0.9\columnwidth}{!}{%
\begin{tabular}{|l|l|l|l|l|}
\hline
        & \multicolumn{2}{c|}{FAM}& \multicolumn{2}{c|}{SSCA}\\ \hline
        & Time & Speedup & Time & Speedup\\ \hline
CPU     & 0.194 s         & 1             & 11.3 s         & 1                   \\ \hline
GPU     & 2.791 ms        & 69.51         & 217 ms         & 52.07               \\ \hline
VCK5000 & 0.630 ms         & 307.94       & 114 ms         & 99.12               \\ \hline
\end{tabular}%
}
\end{table}

\begin{figure}[t]
    \centering
    \includegraphics[width=
    \linewidth]{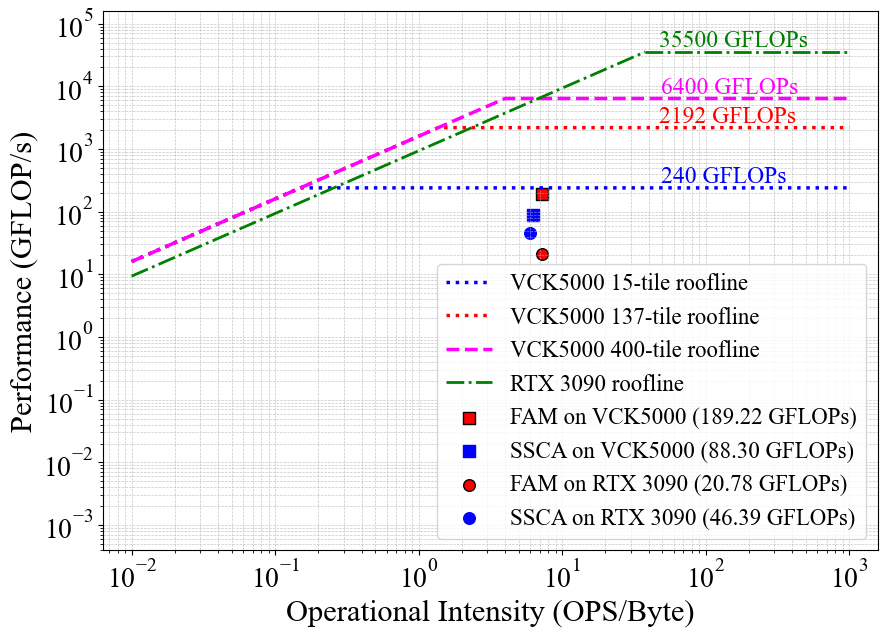}
    \caption{Rooflines of our Implementations.
    }
    \vspace{-2ex}
    \label{fig:roofline}
\end{figure}

The code running on the CPU was compiled using g++ version 9.4.0 with the ``-O2'' optimisation flag. The GPU code was compiled using nvcc, with the host compiler set to g++. The compilation targeted CUDA architecture ``sm\_86'', using C++17 standard with the ``-O3'' optimisation flag. Additionally, the code is linked with the cuFFT library to support efficient FFT operations~\cite{cuFFT2024}. 

In Tab.~\ref{tab:Comparison} we compare our implementation with existing \gls{fpga} designs. In Ref.~\cite{10.1145/3546181}, a quarter \gls{scd} is implemented for comparison with our implementation. To enable a fair comparison with our full \gls{scd}, the reported initiation interval was scaled by a factor of four.

Tab.~\ref{tab:Chap4_speedup1}
presents the execution times (measured on the physical platform) for the \gls{ssca} and \gls{fam} algorithms on CPU, GPU, and VCK5000 platforms. For \gls{fam}, the VCK5000 showed a speedup of 308x compared to CPU and 4.43x over GPU. For \gls{ssca}, the VCK5000 achieved a speedup of 99.12x over the CPU and 1.90x over the GPU.

As shown in the roofline plot of Fig.~\ref{fig:roofline}, the \gls{fam} implementation on VCK5000 achieves 189~GFLOPs, corresponding to 8.6\% of the 137-tile peak performance (2192~GFLOPs). The \gls{ssca} implementation reaches 88.30~GFLOPs, achieving 37\% of its 15-tile peak (240~GFLOPs).
The relatively low utilisation of the \gls{fam} design is attributed to its architectural design: although 137 \gls{aie} tiles are allocated, a significant portion of them are used solely for data movement rather than floating-point operations. This leads to the under-utilisation of available computational resources.
The performance of \gls{ssca} is constrained by the bandwidth between \gls{pl} and \gls{ddrmc}. In our system, communication with 15 \gls{aie} tiles saturates this available bandwidth. With increased off-chip bandwidth, additional \gls{aie} tiles could be utilised to further enhance performance.
On the RTX 3090, both \gls{fam} and \gls{ssca} implementations are memory-bound, achieving 20.78~GFLOPs and 46.39~GFLOPs respectively, also far below the GPU's ceiling of 35~TFLOPs.

We measure power consumption on the VCK5000 and GPU using the ``xbutil'' and ``nvidia-smi'' command-line tools, respectively. For \gls{fam} and \gls{ssca}, the VCK5000 consumes 17~W and 8~W, respectively, on top of an idle power of 23~W. This value is consistent with the power estimate from Xilinx Power Design Manager which reported 13.7~W for FAM. The GPU requires 117~W and 103~W with an idle of 33~W. Consequently, compared to \gls{gpu}, the VCK5000 achieves a 30.5x higher energy efficiency for \gls{fam} and 24.5x higher for \gls{ssca}. Our \gls{ssca} design requires fewer \gls{aie} tiles, resulting in lower power usage on the VCK5000. 


\section{Conclusion}
\label{se:Conclusion}
This paper presents a novel design methodology for high-speed implementations of the \gls{fam} and the \gls{ssca} on the Versal platform. For the \gls{ssca} implementation, we take advantage of the heterogeneous nature of the Versal architecture and utilise the \gls{aie} array’s parallel compute capabilities in parallel with the \gls{pl} to minimise data transfers and manage large intermediate matrices. Our design demonstrates the potential of Versal devices for real-time cyclostationary signal analysis, paving the way for future integration with \gls{sdr} front-ends and machine learning back-ends in advanced \gls{rfml} applications.

\clearpage
\bibliographystyle{IEEEtran}
\bibliography{ref}{}

\begin{thebibliography}{10}
\providecommand{\url}[1]{#1}
\csname url@samestyle\endcsname
\providecommand{\newblock}{\relax}
\providecommand{\bibinfo}[2]{#2}
\providecommand{\BIBentrySTDinterwordspacing}{\spaceskip=0pt\relax}
\providecommand{\BIBentryALTinterwordstretchfactor}{4}
\providecommand{\BIBentryALTinterwordspacing}{\spaceskip=\fontdimen2\font plus
\BIBentryALTinterwordstretchfactor\fontdimen3\font minus \fontdimen4\font\relax}
\providecommand{\BIBforeignlanguage}[2]{{%
\expandafter\ifx\csname l@#1\endcsname\relax
\typeout{** WARNING: IEEEtran.bst: No hyphenation pattern has been}%
\typeout{** loaded for the language `#1'. Using the pattern for}%
\typeout{** the default language instead.}%
\else
\language=\csname l@#1\endcsname
\fi
#2}}
\providecommand{\BIBdecl}{\relax}
\BIBdecl

\bibitem{gardner2006cyclostationarity}
W.~A. Gardner, A.~Napolitano, and L.~Paura, ``Cyclostationarity: Half a century of research,'' \emph{Signal processing}, vol.~86, no.~4, pp. 639--697, 2006.

\bibitem{5067400}
B.~Ramkumar, ``Automatic modulation classification for cognitive radios using cyclic feature detection,'' \emph{IEEE Circuits and Systems Magazine}, vol.~9, no.~2, pp. 27--45, 2009.

\bibitem{9852206}
X.~Liu, C.~J. Li, C.~T. Jin, and P.~H.~W. Leong, ``Wireless signal representation techniques for automatic modulation classification,'' \emph{IEEE Access}, vol.~10, pp. 84\,166--84\,187, 2022.

\bibitem{Gardner91}
W.~A. {Gardner}, ``{Exploitation of spectral redundancy in cyclostationary signals},'' \emph{IEEE Signal Processing Magazine}, vol.~8, pp. 14--36, Apr. 1991.

\bibitem{gardner1986spectral}
W.~A. Gardner, ``The spectral correlation theory of cyclostationary time-series,'' \emph{Signal processing}, vol.~11, no.~1, pp. 13--36, 1986.

\bibitem{amd2024am009}
\BIBentryALTinterwordspacing
{AMD Xilinx}, \emph{AM009 Versal AI Engine}, 2021, versal ACAP AI Engine Architecture Manual. [Online]. Available: \url{https://www.xilinx.com}
\BIBentrySTDinterwordspacing

\bibitem{amd2024xmp452}
\BIBentryALTinterwordspacing
------, \emph{XMP452 Versal AI Core Series Product Selection Guide}, 2024, versal AI Core Series Product Selection Guide. [Online]. Available: \url{https://www.xilinx.com}
\BIBentrySTDinterwordspacing

\bibitem{10.1145/3567429}
\BIBentryALTinterwordspacing
C.~J. Li, X.~Li, B.~Lou, C.~T. Jin, D.~Boland, and P.~H.~W. Leong, ``Fixed-point fpga implementation of the fft accumulation method for real-time cyclostationary analysis,'' \emph{ACM Trans. Reconfigurable Technol. Syst.}, vol.~16, no.~3, Jun. 2023. [Online]. Available: \url{https://doi.org/10.1145/3567429}
\BIBentrySTDinterwordspacing

\bibitem{roberts1991computationally}
R.~S. Roberts, W.~A. Brown, and H.~H. Loomis, ``Computationally efficient algorithms for cyclic spectral analysis,'' \emph{IEEE Signal Processing Magazine}, vol.~8, no.~2, pp. 38--49, 1991.

\bibitem{brown1993digital}
W.~A. Brown and H.~H. Loomis, ``Digital implementations of spectral correlation analyzers,'' \emph{IEEE Transactions on Signal Processing}, vol.~41, no.~2, pp. 703--720, 1993.

\bibitem{gardner1994cyclostationarity}
W.~A. Gardner, \emph{Cyclostationarity in communications and signal processing}.\hskip 1em plus 0.5em minus 0.4em\relax New York: IEEE Press, 1994.

\bibitem{10.1145/3546181}
\BIBentryALTinterwordspacing
X.~Li, D.~L. Maskell, C.~J. Li, P.~H.~W. Leong, and D.~Boland, ``A scalable systolic accelerator for estimation of the spectral correlation density function and its fpga implementation,'' \emph{ACM Trans. Reconfigurable Technol. Syst.}, vol.~16, no.~1, Dec. 2022. [Online]. Available: \url{https://doi.org/10.1145/3546181}
\BIBentrySTDinterwordspacing

\bibitem{Brown1987phdthesis}
W.~A. Brown, ``On the theory of cyclostationary signals,'' Ph.D. dissertation, University of California Davis, 1987.

\bibitem{defence1994implementation}
\BIBentryALTinterwordspacing
E.~April, \emph{On the Implementation of the Strip Spectral Correlation Algorithm for Cyclic Spectrum Estimation}, ser. DREO technical note.\hskip 1em plus 0.5em minus 0.4em\relax Defence Research Establishment Ottawa, 1994. [Online]. Available: \url{https://books.google.com.au/books?id=7QD7MwEACAAJ}
\BIBentrySTDinterwordspacing

\bibitem{netVCK5000}
``Vck5000 versal development card,'' \url{https://www.xilinx.com/products/boards-and-kits/vck5000.html}, accessed: 2025-03-28.

\bibitem{amdUG1076}
\BIBentryALTinterwordspacing
{AMD Xilinx}, \emph{UG1076 Versal ACAP AI Engine Programming Environment User Guide}, 2022, running Software Emulation chapter. [Online]. Available: \url{https://docs.amd.com/r/2022.1-English/ug1076-ai-engine-environment/Running-the-System-in-Hardware}
\BIBentrySTDinterwordspacing

\bibitem{ANTONI2017248}
\BIBentryALTinterwordspacing
J.~Antoni, G.~Xin, and N.~Hamzaoui, ``Fast computation of the spectral correlation,'' \emph{Mechanical Systems and Signal Processing}, vol.~92, pp. 248--277, 2017. [Online]. Available: \url{https://www.sciencedirect.com/science/article/pii/S0888327017300134}
\BIBentrySTDinterwordspacing

\bibitem{10.1111/j.2517-6161.1958.tb00300.x}
\BIBentryALTinterwordspacing
I.~J. Good, ``The interaction algorithm and practical fourier analysis,'' \emph{Journal of the Royal Statistical Society: Series B (Methodological)}, vol.~20, no.~2, pp. 361--372, 12 2018. [Online]. Available: \url{https://doi.org/10.1111/j.2517-6161.1958.tb00300.x}
\BIBentrySTDinterwordspacing

\bibitem{alma991004238809705106}
L.~R. Rabiner and B.~Gold, \emph{\BIBforeignlanguage{eng}{Theory and application of digital signal processing}}.\hskip 1em plus 0.5em minus 0.4em\relax Englewood Cliffs, N.J: Prentice-Hall, 1975.

\bibitem{FFT1M}
\BIBentryALTinterwordspacing
{AMD Xilinx}, ``1 million point float {FFT} @ 32 {Gsps} on {AI Engine},'' 2024, accessed: 2025-03-28. [Online]. Available: \url{https://github.com/Xilinx/Vitis-Tutorials/tree/2024.2/AI_Engine_Development/AIE/Design_Tutorials/16-1M-Point-FFT-32Gsps}
\BIBentrySTDinterwordspacing

\bibitem{8734795}
H.~Kanders, T.~Mellqvist, M.~Garrido, K.~Palmkvist, and O.~Gustafsson, ``A 1 million-point {FFT} on a single fpga,'' \emph{IEEE Transactions on Circuits and Systems I: Regular Papers}, vol.~66, no.~10, pp. 3863--3873, 2019.

\bibitem{cuFFT2024}
``{cuFFT API} reference,'' \url{https://docs.nvidia.com/cuda/cufft/}, accessed: 2025-03-28.

\end{thebibliography}

\end{document}